\documentclass[12pt]{article}

\usepackage{amsbsy}
\usepackage{amssymb}
\usepackage{amsmath}
\makeatletter
\renewcommand\section{\@startsection {section}{1}{\z@}%
                                 {-3.5ex \@plus -1ex \@minus -.2ex}%nn
                                   {2.3ex \@plus.2ex}%
                                   {\normalfont\large\bfseries}}
\renewcommand\subsection{\@startsection{subsection}{2}{\z@}%
                                   {-3.25ex\@plus -1ex \@minus -.2ex}%
                                     {1.5ex \@plus .2ex}%
                                     {\normalfont\bfseries}}
\renewcommand\subsubsection{\@startsection{subsubsection}{3}{\z@}%
                                   {-3.25ex\@plus -1ex \@minus -.2ex}%
                                     {1.5ex \@plus .2ex}%
                                     {\normalfont\itshape}}
\makeatother

\newcommand{\non}{\nonumber \\}

\def\alp{\alpha}   \def\bet{\beta}    
\def\del{\delta}   \def\eps{\epsilon} 
    \def\th{\theta}    
    \def\kap{\kappa}   \def\lam{\lambda}

\def\Gam{\Gamma}   \def\Del{\Delta}   
     
\def\Ome{\Omega}

\def\cM{{\cal M}}  \def\cO{{\cal O}}
\def\cP{{\cal P}}  \def\cR{{\cal R}}

\def\pa{\partial}

\def\inv{^{-1}}

\def\pr{^{\prime}}

\def\rar{\rightarrow}

\def\one{1\!\!1\,\,}

\newcommand{\tr}{\mbox{Tr}}

\def\hlf{\frac{1}{2}}
\def\ove#1{\frac{1}{#1}}

\def\RR{\mathbb{R}}

\def\Str{{\rm Str}}

\def\mtrix#1{\begin{matrix} #1 \end{matrix}}
\newcommand{\str}{\mathrm{Str}}
\newcommand{\dif}{\mathrm{d}}

\begin{document}

\thispagestyle{empty}
\begin{flushright}
\parbox[t]{2in}{CU-TP-1078\\
ITFA-2002-59\\
  hep-th/0212250}
\end{flushright}

\vspace*{0.2in}

\begin{center}
{\Large \bf Evidence for a gravitational Myers effect }

\vspace*{0.3in}
Jan de Boer\footnote{Email address: jdeboer@science.uva.nl},
~Eric Gimon\footnote{Email address: gimon@ias.edu},
~Koenraad Schalm\footnote{Email address: kschalm@phys.columbia.edu}
~and Jeroen Wijnhout\footnote{Email address:
  wijnhout@science.uva.nl}\\[.3in] 

${}^{1,4}${\em Institute for Theoretical Physics \\
  University of Amsterdam \\
  Valckeniersstraat 65 \\
  1018 XE, Amsterdam }\\[.1in]
${}^2${\em School of Natural Sciences\\
Institute for Advanced Studies \\
Princeton, NJ 08540}\\[.1in]
${}^3${\em Department of Physics \\
Columbia University \\
New York, NY 10027}\\[.2in]

{\bf Abstract}
\end{center}

\noindent
An indication for the existence of a collective Myers solution in the
non-abelian D0-brane Born-Infeld action is
the presence of a tachyonic
mode in fluctuations around 
the standard diagonal background. We show that this 
computation for non-abelian D0-branes in curved space has the geometric
interpretation of computing
the eigenvalues of the geodesic deviation operator for $U(N)$-valued
coordinates. On general grounds one therefore expects a geometric
Myers effect in regions of sufficiently negative curvature. We confirm this by
explicit computations for non-abelian D0-branes on a sphere and a
hyperboloid. For the former the diagonal solution is stable, but not
so for the latter. We conclude by showing 
that near the horizon of a Schwarzschild
black hole one also finds a tachyonic mode in the fluctuation
spectrum, signaling the possibility
of a near-horizon 
gravitationally induced Myers
effect.

\newpage
\setcounter{footnote}{0}
\section{Introduction}

Singularities are both the plague and wonder 
of classical general relativity. String theory, however, 
or a
fully quantum theory
of gravity, is expected to be nonsingular.
In
string theory, some gravitational 
singularities are known to be resolved by 
including contributions to the low energy effective action 
of highly massive stringy 
fields or branes which become massless near the classical
singularity (see
e.g. \cite{aspinwallgreenemorrison,witten,stromgreenemor}). 
A second mechanism is that the singularity may be an
artifact of the classical approximation and there are quantum effects
which modify or shield its problematic nature
\cite{polstrass,klebstrass,enhancon}. If the singularity is 
naturally 
associated with a gauge field, the mechanism of shielding/resolving 
is by now 
well-understood. Due to the Myers dielectric effect \cite{Myers}, 
marginally bound strings and/or
other extended objects sensitive to the singularity will polarize to a
 collective higher dimensional extended object. 
This smudges the pointlike
sensitivity, making the singularity effectively invisible.

The Myers effect is a profound discovery and is relevant in many other
situations as well, e.g. it is responsible for the stringy exclusion
principle for so-called giant gravitons \cite{gg} and explains the 
massive supersymmetric membrane worldvolume action in maximally
symmetric pp-wave backgrounds \cite{bmn}. Due to its versatile nature,
it begs the question how general such induced collective behaviour
is in the presence of non-zero background fields. The Myers effect was
originally discovered for non-abelian D0-branes in Ramond-Ramond 
gauge-fields. S-duality allows one to straightforwardly conclude
that Neveu-Schwarz fields can be responsible for such an effect as
well. The question we will address here is whether gravitational
curvature {\em itself} may induce collective behaviour.\footnote{This
  question has been addressed earlier in \cite{Sahakian}. 
These papers investigate 
whether the RR-Myers solution can be molded to a
gravitational Myers solution. The gravitational potential alone is
insufficient to make the specific RR-Myers collective 
D0-brane configuration 
stable, however.} 

We will study this question in 
the same situation where the original Myers
effect was discovered: $N$ marginally bound non-abelian
D0-branes. How to
couple non-abelian D0-branes to gravity, has been a long-standing question. In \cite{DeBoer:2001uk} two
of us
gave an algorithm which imposes the constraints of 
diffeomorphism invariance on 
the action for D0-branes in curved backgrounds. The 
gauge field for diffeomorphisms is the graviton, although the
requirement of general coordinate invariance is not stringent enough
to determine the action uniquely.
This
algorithm implements diffeomorphism invariance in an indirect way as
basepoint invariance between various Riemann normal coordinate
systems. In particular we showed that the often used 
symmetrized trace
approximation is correct to linear
order in the background field, as first found in
\cite{taylorvanraamsdonk}, though it breaks down at the next order. 
Moreover, to this same linear order of
approximation, one may in fact use any coordinate system and one is not
hampered by the normal coordinate requirement. What we will find is
that this approximation, which simplifies the computation
dramatically, is already sufficient to study the presence or absence
of a gravitational Myers effect.

Naturally the details of the final collective state will depend
heavily on the particular choice of gravitational background. This
makes it difficult to explicitly construct a collective
lesser energetic solution which would prove the existence of a
geometrically induced Myers effect. We will follow a different
road. In section 2 we will review how the presence of a
Myers-dielectric solution is associated with the presence of a
tachyonic mode in the off-diagonal 
fluctuation spectrum around the naive
diagonal ``individual'' 
solution. In a curved background the ``individual''
diagonal solution consists of $N$ D0-branes following geodesics 
independently. In section 3 we will show that the spectrum of  
fluctuations around a geodesic solution has a beautiful geometric and
physical interpretation. For a single particle/D0-brane the field
equation for quadratic fluctuations is the geodesic deviation
  equation. It computes the gravitational tidal forces
pulling or pushing objects together or apart. Testing for a
  gravitational 
Myers effect for a collection of D0-branes 
is thus the same as computing the {\em non-abelian} \
geodesic
deviation equation. We perform the test in the weak gravity
approximation 
explicitly in the next section, both
in arbitrary and in Riemann normal coordinates. The latter we discuss 
in
light of its relevance for non-abelian D0-branes coupled to
gravity beyond the linear approximation. Having
set up a general expression for the mass matrix of fluctuations, 
we test 
non-abelian D0-branes on the sphere and the hyperboloid. We find 
that regions
of spacetime with negative curvature have a tachyonic mode in the
 off-diagonal 
fluctuation spectrum. This indicates that there may indeed exist a
purely gravitationally induced Myers effect. To emphasize this
point and test it in a more realistic setting we show that an
off-diagonal 
tachyonic mode is also present near the horizon of a Schwarzschild
black hole. The evidence for
{gravitational} Myers-like behaviour we find here is tantalizing for
its possible consequences and our understanding of string theory as a
theory of quantum gravity. We speculate on this and give an outlook on
future directions in the conclusion, section 5.

\section{The Myers effect: a brief review}

The Myers effect is due to a classical 
instability in the standard
D0-brane configuration in the presence of 
RR-backgrounds. In a flat empty 
background a marginally bound state 
of $N$-coincident D0-branes has as
low energy effective action 
10D $U(N)$ super-Yang-Mills theory reduced
to 0+1 dimensions. Ignoring fermions and 
choosing the gauge $A_0=0$, it reads
\begin{eqnarray}
  \label{eq:22}
  S(X) &=& \frac{m}{2}\int d\tau \tr \dot{X}^i \dot{X}^i
  + \frac{\lam^2}{4}\int 
  d\tau \tr
  \left([X^i,X^j]^2\right) ~,~~~~
  i=1\ldots 
  9 ~.
\end{eqnarray}
The equations of motion have 
a unique static solution comprised by the set of diagonal
matrices 
$X^i={\rm diag}( x^i_\ell)$, with $x^i_{\ell}$ interpreted as the
position of the $\ell$'th D0-brane.\footnote{To derive the low energy 
effective action \eqref{eq:22} from open string theory,
  the appropriate 
solution has $X^i$ proportional to the unit matrix, $X^i =
  x^i\one$. See the second paragraph in the next subsection for a more
detailed discussion of this point.}
 
Turning on RR-fields, specifically a constant IIA RR 4-form,
modifies the D0-brane potential to \cite{Myers}\footnote{The signs and
  factors of 
$i$ follow from requiring reality and positivity; we choose $X^i$ to be
Hermitian.} 
\begin{eqnarray}
  \label{eq:1}
  V(X) = -\frac{\lam^2}{4}\tr \left([X^i,X^j]^2\right)
-\frac{i}{3} F_{0ijk}\tr \left(X^iX^jX^k\dot{X}^0\right)~.
\end{eqnarray}
We will choose the physical gauge $\dot{X}^0 = \one$, 
and we will implicitly limit our attention to the sector $i=1,2,3$, and
ignore the remaining six directions, i.e. we choose 
only $F_{0123} \neq 0$. 

This new term destabilizes the old minima of the potential. 
Eq. \eqref{eq:1} has extrema at
\begin{eqnarray}
  \label{eq:2} 
  \frac{\pa V}{\pa X^i} = 0 = - \lam^2 [X^j,[X^i,X^j]] -\frac{i}{2}
F_{0ijk}[X^j,X^k]~.
\end{eqnarray}
The potential has ``two''
qualitatively different minima, corresponding to static solutions to
the field equations.  The first is the standard solution with $X^i$
  diagonal, i.e. $N$ independent D0-branes, 
and vanishing rest energy. 
The second set of solutions consists of  
non-abelian solutions 
$[X^i,X^j]=-\frac{i}{2\lam^2} F_{0ijk}X^k$ with negative rest energy
$\lam^2 E_{rest}=(\ove{16}-
\ove{12})(F_{0ijk}F_0^{~ijm}\tr(X^kX_m))$. Here the internal
confining force from the commutator squared potential, $[X^i,X^j]^2$, 
  is balanced by the externally induced dielectric 
stretching. Energetically these second solutions are favoured: 
a cluster of unattached
D0-branes let loose, prefers to form a collective dielectric state,
  rather then each pursuing its own course.

\subsection{Evidence for the Myers effect: quadratic fluctuations}

The D0-brane configuration described by the 
second non-abelian solution had already appeared in the
literature prior to Myers' discovery of the 
polarization term in \eqref{eq:1} \cite{Kabat:1997im}. The
polarization term, however, is crucial in ensuring that the solution
is energetically favoured.
Suppose that one didn't know that such a stable non-abelian 
Myers solution existed. Before
tackling the difficult problem of finding new lesser energetic solutions to the field
equations by brute force, one could deduce that such a solution
exists by calculating the spectrum of quadratic fluctuations
around the simple commuting background solution. The presence of the
second less energetic type of
solutions is reflected in the presence of an 
tachyonic instability in the off-diagonal fluctuations.
\cite{Jatkar}. Vice versa, if a tachyonic mode
is present, this signals, bar run-away behaviour, a non-abelian
collective mode solution.
For non-abelian D0-branes coupled to gravity this is the road we shall
pursue, rather than attempt to find an explicit less energetic
solution to the field equations directly.

As a warm-up exercise, let us confirm this
philosophy for the original Myers effect. We will calculate the spectrum of
quadratic fluctuations for two flat-space D0-branes coupled to a
constant RR-four-form background, as in
\cite{Jatkar}. We will do
this starting from the field equation (\ref{eq:2}), rather
than the action (\ref{eq:1}). This way we directly obtain the forces
experienced  by the D0-branes and it will be a convenient point
of view when we discuss D0-branes coupled to gravity
in the next section.   

We should
make one remark at the outset. Already for flat space non-abelian
D0-branes coupled to RR-fields, the space of
solutions and spectrum of fluctuations 
is very complicated \cite{Jatkar}. In fact, the 
strict low energy background for a cluster of superposed D0-branes,
where the fields $X^i$ are proportional to the unit matrix,
$X^i=x^i\one$,
does {\em not} have a tachyonic instability which 
signals the presence of the dielectric solution. 
This mode is only visible, as we will see, when the diagonal
entries of the background solution are different. This is, strictly
speaking, in conflict with the use of the 
non-abelian effective action in eq. \eqref{eq:22}, 
for off-diagonal modes are massive in this
case. However,
as long as the masses proportional to the ``distance'' between the
diagonal entries are less than the string 
length, the action should be a good approximation. 
Specifically the regime where we may trust the action is when the
following inequalities for 
$X^i$, $m_{D0}=1/g_s\ell_s$, and $\lam^2 =1/g_s\ell_s^5$ hold 
\cite{dkps}
\begin{eqnarray}
  \label{eq:3a}
   \ell_s\left(\frac{d}{d\tau}\right)^{n+1} X^i &\ll& 
   \left(\frac{d}{d\tau}\right)^{n} X^i ~~~\forall~ n \geq 1 ~,\\
\label{eq:3b}
   \frac{d}{d\tau} X^i &\ll& 1~,
\end{eqnarray}
and
\begin{eqnarray}
\label{eq:3c}  
\hspace{-1in}
(\lam\sqrt{m})^{-1/3} = g_s^{1/3}\ell_s & \ll & X^i ~~\ll 
~~\ell_s 
~.
\end{eqnarray}
The first equation simply states the
regime 
of validity for the Born-Infeld action: fields must be slowly varying. 
The second allows the
truncation to second order in velocities. The r.h.s. of the third
inequality is also a validity bound of the Born-Infeld 
action: the Born-Infeld action is a Wilsonian effective action with
energies and masses below the string scale
$\ell_s\inv=(\alp\pr)^{-1/2}$.  
The left-hand side is the bound set by the eleven dimensional Planck
length $\ell_{11}=g_s^{1/3}\ell_s$. 
In pure field theory terms, it is the bound
which 
validates the use of perturbation theory.

\subsubsection{Computation of the spectrum of fluctuations}

Consider thus a small
variation $\del X^i$ around a diagonal background solution
$\tilde{X}^i$. The change in the field equation is 
\begin{eqnarray}
  \label{eq:6}
  \left.\del\frac{\pa V}{\pa X^i}\right|_{X^i=\tilde{X}^i} &=&
  \left. -iF_{0ijk}[\del X^j,\tilde{X}^k] - \lam^2 [\tilde{X}^j,[\del
  X^i,\tilde{X}^j]]- \lam^2[\tilde{X}^j,[\tilde{X}^i,\del
  X^j]]\right|_{X^i=\tilde{X}^i}~. 
\end{eqnarray}
Denoting entries of $\tilde{X}^i$ and the fluctuations $\del X^i$ as
\begin{eqnarray}
  \label{eq:7}
  \tilde{X}^i =\left(
  \begin{matrix}
   x_1^i & 0\cr 0 & x_2^i
  \end{matrix}
  \right)~,~~ \del
  X^i = \left(
  \begin{matrix}
  a^i & b^i \cr \bar{b}^i & d^i
  \end{matrix}
  \right) ~,~~~a,d
  \in \RR^d~,
\end{eqnarray}
the ``elementary'' commutator $[\del X^i, \tilde{X}^j]$ equals
\begin{eqnarray}
  \label{eq:8}
  [\del X^i, \tilde{X}^j] = \left(\mtrix{ 0 & b^i(x_2-x_1)^j \cr \bar{b}^i(x_1-x_2)^j & 0 } \right)~.
\end{eqnarray}
The Gauss law $[\dot{X}^i,X^i]=0$ tells us that only fluctuations
  transverse to the static diagonal background are 
  dynamical. Physical fluctuations therefore satisfy $[\del
  X^i,\tilde{X}^i] = 0$.  Substituting we find 
$\sum_i b^i(x_1-x_2)^i= 0$, i.e. physical fluctuations $b^i$ are
  orthogonal to the D0-brane separation $\Del^i \equiv (x_1^i-x_2^i)$. 
This implies that the last term in (\ref{eq:6}) vanishes. The
remaining terms are easily computed to be 
\begin{eqnarray}
  \label{eq:9}
  \left.\del\frac{\pa V}{\pa X^i}\right|_{X^i=\tilde{X}^i} &=&
  \left(\mtrix{ 0 & i(b\times \Del)^i + b^i\lam^2\Del^2 \cr -i(\bar{b}\times \Del)^i+\bar{b}^i \lam^2\Del^2 & 0}\right)~.
\end{eqnarray}
We have defined $(a \times b)^i = F_{0ijk} a^j b^k$. 
The quadratic fluctuation
matrix is by definition
\begin{eqnarray}
  \label{eq:10}
  \left.\del\frac{\pa V}{\pa X^{ib}}\right|_{X^i=\tilde{X}^i} &=& M_{ib,ja} \del
  X^{j a} ~~~~~,~ a=1,...,N^2~,
\end{eqnarray}
where we consider the matrix fields $X^i$ as column vectors. Rewriting
eq. (\ref{eq:9}) as such a mass term, we find 
\begin{eqnarray}
  \label{eq:11}
  \left.\del^2 V \right|_{X^i=\tilde{X}^i} &=& \int d\tau ~
  b^i\left(\del_{ij}\lam^2\Del^2 + i F_{0ikj}\Del^k \right)\bar{b}^j + {\rm c.c.}
\end{eqnarray}
The masses are then given by the eigenvalues of this complex $d\times d=3\times 3$ matrix. 
Setting $F_{0ijk} = \lam^2 \rho \eps_{ijk}$ and recalling that $i,j,k$ take
values in the range $i=1,2,3$, one finds the 
characteristic polynomial
\begin{eqnarray}
  \label{eq:13}
  \left| (-\xi+\Del^2)\one_{3\times 3}+i\rho \eps_{ikj}\Del^k \right|
  = (-\xi+\Del^2)^3-\rho^2(-\xi+\Del^2)\Del^2=0 ~,
\end{eqnarray}
with solutions\footnote{This set of
solutions eq. (\ref{eq:14}) corresponds to Appendix A of
\cite{Jatkar}; The three remaining zero modes in the solution
presented there are
the freedom to set the values $a^i$ in $\del X^i$ of
eq. (\ref{eq:7}); for traceless fluctuations as they consider $d^i$
then equals $d^i=-a^i$.}
\begin{eqnarray}
  \label{eq:14}
  \xi &=& \Del^2~,~~~~~~~~~~~~~~~~~~{\rm degeneracy:2},\non
 (-\xi+\Del^2)^2-\rho^2\Del^2=0 ~~\Rightarrow~
~\xi &=& \Del^2 \pm \rho \Del
 ~,~~~~~~~~~~{\rm degeneracy:2 (each)}.
\end{eqnarray}
For small values of $\Del$ and $\rho$ of order unity, 
the second set of solutions will have negative eigenvalues which
signal the presence of tachyons. These are responsible for the Myers effect. Note that for $\rho=0$, we recover
that the off-diagonal fluctuations are proportional to the distance between the D0-branes. But also note how for
$\Del=0$ {\em all} tachyonic modes are absent. 
As we forewarned, truly superposed D0-branes are in fact marginally
stable. If we insist on the validity of 
perturbation theory, though, $\Del$ has a minimal length 
set by eq. \eqref{eq:3c}. In that case $\
Del$ strictly vanishing is not allowed, and the diagonal solution is unstable.

In summary, turning on an RR-background field exerts a destabilizing
force on the naive commuting solution to the D0-brane field
equations. The signal of destabilization is the presence of tachyonic
modes in the fluctuation spectra of the off-diagonal degrees of
freedom. The presence of these modes indicates that there exists a
less energetic solution to the field equations, where the internal
confining non-abelian potential forces are balanced by the external
dielectric repulsing forces. Physically the commuting solution
corresponds to a background of $N$ independent D0-branes, whereas in 
the non-abelian solution the D0-branes behave collectively as an
extended object, a D2-brane.

\section{D0-branes and Gravity}
\setcounter{equation}{0}

The question we seek to answer here, is whether the
  gravitational force can have a similar destabilizing influence on
  the commuting background solution. Bar runaway solutions, 
this would indicate the existence of a similar non-abelian mode, where
the $N$ D0-branes behave collectively as an extended object. 

To investigate this we need to know the action for
non-abelian D0-branes in curved space. 
How to couple the non-abelian D0-brane action to gravity has been a
prominent question since their discovery. The fields $X^i$
describing the transverse coordinates are now $U(N)$-valued and no
longer commute. Any term in the action beyond quadratic order
needs a specific ordering instruction. In particular the naive
introduction of a metric into the kinetic term of the action \eqref{eq:22},
\begin{eqnarray}
  \label{eq:45}
  S_{naive,kin} \sim \frac{m}{2} \int d\tau \tr G_{ij}(X)\dot{X}^i\dot{X}^j~,
\end{eqnarray}
is ambiguous. In principle the action, and therefore the ordering, is
derivable from string-theory amplitude-computations \cite{o}. 
In practice this
is a daunting task, and the general approach has been to construct the
action {\em ab initio} subject to a set of consistency conditions and
constraints \cite{dko}. 
Based on the general
properties of such amplitudes, and known physical quantities
which the correctly ordered action should reproduce, Douglas put forth
forth a set of axioms, which the action for D0-branes in curved space
ought to obey \cite{Douglas}. Concretely they are
\begin{itemize}
\item The action must contain a single $U(N)$ trace \cite{tseytlin}.
\item For diagonal matrices the action must reduce to $N$ copies of
  the first quantized particle action in curved space.
\item The moduli-space of static solutions to the field equations must
  equal $N$-copies of the (spacetime) manifold $M$ modulo the action
  of the permutation group: $\cM =M^N/S_N$.
\end{itemize}
and
\begin{itemize}
\item {\em Masses of off-diagonal fluctuations around a diagonal
    background $X^i={\rm diag} (x^i_\lam)$, 
should be proportional to the geodesic distance between the
corresponding entries.}
\end{itemize}
Collectively these requirements are known as the axioms of
D-geometry. 

In
\cite{DeBoer:2001uk} two of us derived an algorithm to constrain the
ordering of the action
using a more fundamental principle. 
The characteristic feature of {\em any} action
coupled to gravity is diffeomorphism invariance. The presence of some
form of coordinate invariance for matrix-valued coordinates, together
with the single trace requirement and that for linearized gravity the
ordering would be completely symmetrized (see section 4), 
were the sole input we used to
determine the ordering. The answer does not, surprisingly, appear to
be unique
despite the very stringent nature of the non-linear diffeomorphism
constraints. However, due the fundamental
nature of diffeomorphism invariance, an action obtained by this
algorithm obeys all the axioms of D-geometry.

These very axioms, however, put us in a bind where the question of a
gravitational Myers effect is concerned. In the previous section we
showed that the presence of the Myers effect is signaled by tachyonic
modes in the quadratic fluctuation spectrum. But
the fourth, italicized, axiom of D-geometry 
states that the mass-matrix of off-diagonal
fluctuations should be proportional 
to geodesic lengths, which are ``strictly positive''. Hence one's
initial reaction is  
to say that no
gravitational Myers 
effect exists. However, careful consideration of the D-geometry axioms
shows that the fourth axiom
is only applicable to infinitesimal fluctuations around a {\em static} solution to the
field equations.

Static solutions are not very natural in a curved
background. A first consequence of the
gravitational curvature is the exertion of a gravitational force on the
cluster of D0-branes, accelerating them away from zero
velocity. Generically we expect that each of the D0-branes
will independently follow a geodesic. 
In fact, D-geometry requires that
the naive solution for a cluster of
D0-branes, each following geodesic flow independently, 
remains a solution to the non-abelian field equations. 
It is in the perturbation around such a 
set of independent geodesics rather than the static 
solution that we
should search for negative eigenvalues in the mass matrix of off-diagonal
fluctuations.
The D0-branes must therefore have nonzero velocity for the collective
behaviour to occur. 
Hence a Myers effect 
arises from balancing a stretching induced by the, 
now non-zero, kinetic term and the
confining potential term. This in fact 
agrees nicely with the RR-background story, in that
for gravity the kinetic term is the universal coupling. 

And at 
this moment we have arrived on familiar territory. The ``physical''
stretching of an extended object in the presence of gravitational
curvature is a consequence the gravitational 
tidal forces. These forces are
traditionally derived using what is known as the geodesic deviation
equation. Perhaps less known is that the geodesic deviation equation
can be derived from the particle action in curved space by
considering quadratic fluctuations around a geodesic solution
\cite{vh}.
This puts the procedure outlined above on firm geometric
footing. {\em Testing for the Myers effect is equivalent to computing
non-abelian geodesic deviation.}

\subsection{Quadratic fluctuations, geodesic deviation and tidal forces}

Let us briefly recall 
this connection between quadratic fluctuations and the geodesic
deviation equation for the standard particle case (see e.g. \cite{vh}). 
The action is the proper distance,
\begin{equation}
  \label{eq:12}
  S_{part} = m \int d\tau \sqrt{ -g_{ij} \dot{x}^i \dot{x}^j}~,
\end{equation}
with as field equation the geodesic equation
\begin{equation}
  \label{eq:15}
  \frac{\pa S}{\pa x^i} =0= g_{ij}\ddot{x}^j +
  \Gam_{ijk}\dot{x}^j\dot{x}^k  = g_{ij} \dot{x}^k \nabla_k \dot{x}^j \equiv g_{ij}\nabla_{\tau} \dot{x}^j~.
\end{equation}
Consider now the quantity $\nabla_{\tau} \dot{x}^j$ and vary its dependence on $x^i$ one more time. This can be
implemented explicitly by letting $x^i$ depend on some additional ``affine'' parameter $s$,
 vary $\nabla_{\tau} \dot{x}^j$ w.r.t. this parameter $s$, and then extract
an overall
\def\xprim{{x^{\!\prime}}}
factor $\xprim^j =\frac{\pa}{\pa s} x^j$. Of course the resulting combination,
\begin{equation}
  \label{eq:16}
  \delta_s \nabla_{\tau} \dot{x}^j \equiv  \xprim^k \frac{\delta}{\delta x^k}
\nabla_{\tau} \dot{x}^j~,
\end{equation} is not covariant, but can be made so by adding an improvement term,
\begin{equation}
  \label{eq:17}
  \delta_s \nabla_{\tau} \dot{x}^j + \xprim^k\Gam^j_{kn} \nabla_{\tau}
  \dot{x}^n \equiv \nabla_s\nabla_{\tau} \dot{x}^j~.
\end{equation}
Note, however, that the improvement term is directly proportional to the geodesic equation. Therefore, if we
substitute as background value for $x^j$ a solution to the field
equation, the improvement term is formally zero and we may add it for free. As a
consequence
\begin{equation}
  \label{eq:18}
  \left. \delta_s \nabla_{\tau} \dot{x}^j \right|_{x^j~geodesic} =
  \left. \nabla_s\nabla_{\tau} \dot{x}^j \right|_{x^j ~geodesic}~.
\end{equation}
From elementary algebra it follows that\footnote{We are
  using conventions where $R_{\mu\nu\rho\tau} =
  -\pa_{\mu}\Gam_{\tau~\nu\rho} + \Gam^{\kap}_{~\mu\tau}
  \Gam_{\kap~\nu\rho} - (\mu \leftrightarrow \nu)$.}
\begin{eqnarray}
  \label{eq:19}
  \nabla_s\nabla_{\tau} \dot{x}^j &=& R_{s\tau ~k}^{~~j} \dot{x}^k
  +\nabla_{\tau}\nabla_{s} \dot{x}^j \non
&=& R_{s\tau ~k}^{~~j} \dot{x}^k
  +\nabla_{\tau}\nabla_{\tau} \xprim^j~.
\end{eqnarray}
If we therefore start from the geodesic equation,
\begin{equation}
  \label{eq:20}
  \nabla_{\tau}\dot{x}^j = 0~.
\end{equation}
and vary both sides, we find using eq. (\ref{eq:18}) that the r.h.s. of (\ref{eq:19}) must vanish. This is the {\em
geodesic
  deviation 
equation}. It determines the acceleration ($\nabla_{\tau}^2$) of an
infinitesimally close geodesic ($\xprim^j$) in 
terms of the background Riemann tensor. It therefore determines 
how fast infinitesimally close non-interacting particles are pulled
apart/together 
in a gravitational field due to tidal forces.

It is now clear how quadratic fluctuations and the tidal forces are related. The quadratic fluctuation matrix
\begin{eqnarray}
  \label{eq:21}
  \delta \frac{\pa S}{\pa x^i} = M_{ij} \delta x^j~~~~
  \Leftrightarrow ~~~~\delta_s \frac{\pa S}{\pa x^i} &=&
  M_{ij} \xprim^j  \non
&=& (R_{i k j \ell}\dot{x}^k\dot{x}^{\ell} - \nabla_{\tau}\nabla_{\tau}) \xprim^j
 \end{eqnarray}
is the geodesic deviation ``operator''. The geodesic deviation
equation thus computes the kernel of this
operator. Hence, one way to interpret a solution $\del
x^i_{sol}(\tau)$ to the
geodesic deviation equation, is that 
to first order $x^i(\tau)_{geodesic}+\del
x_{sol}^i(\tau)$ 
is then also a solution to the geodesic equation (see
\cite{vh}). 
Since the geodesic equation is the field-equation of
the proper-distance action, this specific 
variation $\del x^i_{sol}(\tau)$, 
such that the geodesic equation is still satisfied,
must be a zero mode in the spectrum of fluctuations determined by the
action. And this is just what eq. (\ref{eq:21}) says.

If the background D0-brane configuration is constant, i.e.
  $\dot{x}^i=0$, the ``geodesic deviation operator'' 
reduces to $\pa/\pa_{\tau}^2$. This is
  the static solution situation appropriate for D-geometry, as
  discussed above. If the D0-branes have non-zero velocity, however,
  the ``geodesic deviation operator'' includes the characteristic 
extra term
  proportional to the Riemann tensor. This term 
  encodes the gravitational potential.  In field theory terms 
  we can interpret it as a mass
  term for the fluctuations. In particular positive mass fluctuations
  correspond to converging geodesics, and negative mass fluctuations
  correspond to ``unstable'' diverging geodesics. For non-abelian
  D0-branes following independent geodesics, these diagonal 
fluctuations will still be present. New, however, will be the presence
  of off-diagonal fluctuations, probing the stability of the
  configuration.    
If there exists a gravitational Myers effect for
  non-abelian D0-branes, it is the non-commutative analog of the extra
  curvature term in the geodesic deviation
operator eq. (\ref{eq:21}) which should be responsible for the destabilization.

\subsubsection{Intuitive instability of D0-branes in a curved
  background} 

The computation that should establish the existence of a gravitational
Myers effect thus fits beautifully in a geometric framework from the
D0-brane perspective. At a fundamental level, D-brane dynamics is
governed by open strings; the D-brane is a (dynamical) defect on which
open strings can end. The RR-Myers effect has a simple physical
interpretation from this point of view. The RR-field polarizes and
stretches the charges at the endpoints of the dynamical open string,
until balanced by the internal string tension. For gravitational
tidal forces one can develop a similar intuitive picture, which
argues that in particular backgrounds a gravitational Myers effect 
  should occur. We can model the string by an elastic cord --- a
massless 
ideal spring --- connecting two infinitesimally separated masses. If
the tidal forces, measured by the geodesic deviation equation are
attractive, the additional presence of the connecting spring will be
qualitatively irrelevant. If the tidal forces are repulsive, however,
one expects the naive geodesic system to be unstable. The preferred
state would be that where the separation is such that the tidal and
spring forces precisely balance each other. 
 Such a toy model can be described by the equation
  \begin{eqnarray}
    \label{eq:46}
    \frac{d^2}{d\tau^2} d(x(\tau),y(\tau))+d(x(\tau),y(\tau))=0~,
  \end{eqnarray}
where $d(x,y)$ is the geodesic distance between the positions of the
two masses, $x$ and $y$. This is an effective equation valid in the
frame of a specific observer, whose clock measures $\tau$. 
Tidal forces are repulsive if the spacetime is negatively
curved. Qualitatively we therefore expect negative curvature to be a
precursor of the gravitational Myers effect. 
The exact computation, with which we will now proceed, will
indeed bear this out.

\section{The geometric Myers effect}
\setcounter{equation}{0}

As we stated in the introduction,
a consistent, though not unique, coupling of non-abelian D0-branes 
to gravity can in principle be algorithmically derived to 
any order of precision  
\cite{DeBoer:2001uk}. The principle 
behind this algorithm is the
requirement of diffeomorphism invariance of the non-abelian
D0-brane action. 
The
algorithm implements diffeomorphism invariance in a non-standard way
as base-point independence between normal coordinate systems
rooted on an arbitrary basepoint $\cP$. This paradoxical appearance of
preferential coordinates, however, is only necessary if one goes
beyond linear order in the curvature. At linear order, the non-abelian
nature of the coordinates is only marginally relevant. Within this
approximation one may in fact use any coordinate system, as the
action, specified with the completely symmetrized ordering, 
is up to the same linear order of approximation invariant under
diffeomorphisms \cite{DeBoer:2001uk}.

\subsection{Weak curvature: The symmetrized trace approximation}

A natural first attempt would then 
be to conjecture that a Myers-like
effect is already visible in this simplified situation, i.e. 
when the
spacetime is only slightly distorted and we may consider 
the curvature as a perturbation. 

As we just explained 
the advantage of treating the curvature perturbatively is that to first order in a perturbation the action in
{\em any} coordinate
  system is simply the linearized symmetrized trace action 
action, first found by 
Taylor and van Raamsdonk \cite{taylorvanraamsdonk} and confirmed by
explicit string amplitude calculations \cite{o}. 
\begin{eqnarray}
  \label{eq:24}
  S_{\cO(h)} = \int d\tau \frac{m}{2}\left(
  \eta_{\mu\nu} \tr \dot{X}^{\mu} \dot{X}^{\nu}  +
  \Str (h_{\mu\nu}(X)   \dot{X}^{\mu} \dot{X}^{\nu}) \right)
  +\frac{\lam^2}{2} \Str(h_{\mu\nu}(X)\eta_{\alp\bet}
  [X^{\mu},X^{\alp}][X^{\nu},X^{\bet}]) +\ldots
\end{eqnarray}
This is because the action is invariant under linearized
diffeomorphisms, which is easily verified. 
Once we study specific examples, we will 
therefore be able to avoid the difficult step
of having to transform to normal coordinates first. The restriction to
weak gravity does create an additional bound on the validity of the
action in addition to eqs. \eqref{eq:3a}-\eqref{eq:3c}: the moments of
deviation from the flat metric, including the zeroth one,  
which act as coupling constants in the action
\eqref{eq:24}, must be small
\begin{eqnarray}
  \label{eq:47}
  \left(\ell_s \pa_{y^\alp}\right)^n h_{\mu\nu}(y)|_{y=0}  \ll
  1~~~\forall ~n \geq 0.
\end{eqnarray}
However, we do not need that $\ell_s \pa^{n+1} h_{\mu\nu} \ll \pa^n h_{\mu\nu}$.

We are now
in a position to compute the mass matrix of quadratic fluctuations
around a diagonal background of 
geodesics. There will be two
contributions to the mass matrix, one from the new curvature term in
the geodesic deviation operator which originates in the kinetic term,
and one from the potential term. Let us focus on the new curvature
contribution from the 
kinetic term first.

\subsubsection{The kinetic term}

A convenient way to write the symmetrized trace expression for the
action is by exponentiation
\begin{eqnarray}
  \label{eq:25}
  S_{kin}= \left.\frac{m}{2} 
\int d\tau ~\tr ~e^{\int d\tau\pr X(\tau\pr)
  \cdot \pa_{y(\tau\pr)}}
  g_{\mu\nu}(y)\dot{y}^{\mu}\dot{y}^{\nu}\right|_{y=0} ~.
\end{eqnarray}
where, implicitly, we limit our attention to linear terms in $g_{\mu\nu}(y)=\eta_{\mu\nu}+h_{\mu\nu}(y)$.
To evaluate the exponential, one uses standard functional differentiation,
\begin{eqnarray}
  \label{eq:26}
  \pa_{y^{\nu}(\tau\pr)} y^{\mu}(\tau) =
  \del^{\mu}_{\nu}\del(\tau-\tau\pr)~.
\end{eqnarray}
Varying the fields $X^i$ twice, the formal expression for quadratic
fluctuations around a background solution equals
\begin{eqnarray}
  \label{eq:27}
  \del^2 S_{kin} = \frac{m}{2}  \int d\tau \int_0^1 ds~ \tr ~e^{s\int X \cdot
  \pa_y} \del 
  X \cdot \pa_y \, e^{(1-s) \int X \cdot \pa_y} \del X \cdot \pa_y\,
  g_{\mu\nu}(y)\dot{y}^{\mu}\dot{y}^{\nu}~.
\end{eqnarray}

To evaluate the derivatives we use the identity
\begin{eqnarray}
  \label{eq:28}
  \pa^n_y \pa^m_y f(y) = \left.\pa^n_z\pa^m_y f(y+z)\right|_{z=0}
\end{eqnarray}
to separate derivatives 
contracted with $\del X^{\mu}$ from those in the exponentials.
\begin{eqnarray}
  \label{eq:29}
  \del^2 S_{kin} = \left.\frac{m}{2} \int d\tau \int_0^1 ds~ 
\tr ~e^{s\int X \cdot
  \pa_y} \del 
  X \cdot \pa_z \, e^{(1-s) \int X \cdot \pa_y} \del X \cdot \pa_z\,
  \left
  [ g_{\mu\nu}(y+z)(\dot{y}+\dot{z})^{\mu}
  (\dot{y}+\dot{z})^{\nu}\right]\right|_{z=0} ~.  
\end{eqnarray}
Expanding the argument
$g_{\mu\nu}(y+z)(\dot{y}+\dot{z})^{\mu}(\dot{y}+\dot{z})^{\nu}$  
to second order is by now a well 
known exercise. Using the field equation and following the steps 
leading to eq. (\ref{eq:21}) we
obtain the kernel of the geodesic deviation 
operator
\begin{eqnarray}
  \label{eq:30}
  \del^2 S_{kin} &=& \frac{m}{2} \int d\tau \int_0^1 ds~ 
\tr ~e^{s\int X \cdot
  \pa_y} \del 
  X \cdot \pa_z \, e^{(1-s) \int X \cdot \pa_y} \del X \cdot \pa_z\,
  \left[R_{\mu\alp\bet\nu}(y)\dot{y}^{\alp}\dot{y}^{\bet}z^{\mu}z^{\nu}
  + \right. \non
&& \hspace{3.5in} \left. + g_{\mu\nu}(y)\nabla_{\tau}(y)
  z^{\mu}\nabla_{\tau}(y) z^{\nu} \right]~.
\end{eqnarray} 
The
second term is the kinetic term for the {\em
  fluctuations}. The Christoffel symbols present in the derivatives
serve to covariantize the momenta but do not contribute to the
mass-matrix. This is evident from a computation in normal coordinates,
which will be presented in the next subsection. 
As presaged above, it is the first term, 
proportional to the Riemann
  tensor, which is relevant. Concentrating only on this term,
\begin{eqnarray}
  \label{eq:31}
  \del^2 S_{kin,R} &=& \frac{m}{2} \int d\tau \int_0^1 ds~ \tr~e^{s\int X
  \cdot \pa_y} \del 
  X \cdot \pa_z \, e^{(1-s) \int X \cdot \pa_y} \del X \cdot \pa_z\,
  \left[R_{\mu\alp\bet\nu}(y)
  \dot{y}^{\alp}\dot{y}^{\bet}z^{\mu}z^{\nu}\right]  
  \non
&=& m  \int d\tau \int_0^1 ds~ \tr~e^{s\int X \cdot \pa_y} \del
  X^{\mu}\, e^{(1-s) \int X \cdot \pa_y} \del X^{\nu} \,
  \left[\cR_{\mu\nu}(y)\right]
\end{eqnarray}
with $\cR_{\mu\nu}(y) =
R_{\mu\alp\bet\nu}(y)\dot{y}^{\alp}\dot{y}^{\bet}$, we can simplify it
further. We are 
considering fluctuations around a background $X_{bg}^{\mu}(\tau) =
\mbox{diag}(x_{\ell}^{\mu}(\tau))$ with each 
$x^{\mu}_{\ell}(\tau)$ 
obeying the standard geodesic equation (\ref{eq:15}).
This means that each exponential is in fact
a diagonal matrix. Consider for simplicity a two by two (sub)system of
D0-branes, and parameterize the exponential and the 
fluctuations as
\begin{eqnarray}
  \label{eq:32}
  e^{s\int X \cdot \pa_y} &=& \left(\mtrix{ e^{s\int x_1 \cdot \pa_y}&
                                          0 \cr 
                                          0 & e^{s\int x_2 \cdot
                                          \pa_y}}\right)\equiv
\left(\mtrix{ f_1(s)& 0 \cr
                                          0 & f_2(s)}\right)~, \non
\del X^{\mu} &=& \left(\mtrix{ a^{\mu} & b^{\mu} \cr \bar{b}^{\mu} &
                                          d^{\mu}}\right)~.
\end{eqnarray}
Multiplying and taking the trace of these explicit matrices we find
that 
\begin{eqnarray}
  \label{eq:33}
 \del^2 S_{kin,R} &=& 
 m \int d\tau \int_0^1 ds\left[
f_1(s)f_1(1-s)a^{\mu}a^{\nu} +f_2(s)f_2(1-s)d^{\mu}{d}^{\nu} +
                                          \right. \\ \nonumber
&& \hspace{2in} \left. f_1(s)f_2(1-s)b^{\mu}\bar{b}^{\nu}+
f_2(s)f_1(1-s)\bar{b}^{\mu}b^{\nu}\right]\cR_{\mu\nu}(y)~.
\end{eqnarray}
As $f_i(s)f_i(1-s)=f_i(1)=e^{\int X\cdot \pa_y}$ we immediately recognize the $a^{\mu}a^{\nu}$ term and the
$d^{\mu}d^{\nu}$ term as the  diagonal 
quadratic fluctuations around the $x_1$ and $x_2$ geodesic
respectively. These measure the standard 
(abelian) geodesic deviation. The
parts of interest to us are the new contributions from the off-diagonal
fluctuations $b^i$; these are relevant for the stability of the
configuration. As their mass does not depend on $a^i$ and $d^i$, we
will set $a^i=0=d^i$ in the remainder. 

Noting that  
\begin{eqnarray}
  \label{eq:34}
  f_1(s)f_2(1-s) = e^{\int x_2 \cdot \pa_y + s \int (x_1-x_2) \cdot
  \pa_y } ~,
\end{eqnarray}
we can evaluate all the partial derivatives to 
\begin{eqnarray}
  \label{eq:35}
  f_1(s)f_2(1-s)\cR_{\mu\nu}(y) = \cR_{\mu\nu}(x_2+
  s\Del)~,~~~~\Del^i\equiv (x_1^i-x_2^i)~.
\end{eqnarray}
As final expression for the contribution to the mass matrix of
off-diagonal fluctuations from the kinetic term we obtain
\begin{eqnarray}
  \label{eq:36}
  \del^2 S_{kin,R} &=&  m \int d\tau \int_0^1 ds\left
  [ b^{\mu}\bar{b}^{\nu} 
  \left(\cR_{\mu\nu}(x_2+s\Del)+\cR_{\nu\mu}(x_1-s\Del)\right)\right]~.
\end{eqnarray}
Note that the expression 
looks remarkably like an averaging of the curvature over a
string worldsheet.

To eq. \eqref{eq:36} 
we have to add the contribution from the potential term. 

\subsubsection{The potential term}

To find the quadratic fluctuations of the potential term,
\begin{eqnarray}
  \label{eq:37}
  S_{pot} &=& \frac{\lam^2}{4} \int d\tau \int_0^1 ds~ \tr ~e^{s\int X
  \cdot \pa_y} 
  [X^{\mu},X^{\alp}]e^{(1-s)\int X \cdot \pa_y}
  [X^{\nu},X^{\bet}]g_{\mu\nu}(y) 
  g_{\alp\bet}(y)~,
\end{eqnarray}
(again implicitly limiting our attention to only
the linear term in $h_{\mu\nu}(y)$) is quite a bit easier.  
This is because the commutators vanish when
evaluated on the diagonal background. So a non-vanishing contribution
to the quadratic fluctuations can only occur when we vary
one of the two $X^{\mu}$'s in each of the commutators,
\begin{eqnarray}
  \label{eq:38}
\nonumber \del^2 S_{pot} &=&  \lam^2  \int d\tau \int_0^1 ds
~\tr~ e^{s \int X \cdot \pa_y} \left([\del 
X^{\mu},X^{\alp}]+ [X^{\mu},\del X^{\alp}]\right) e^{(1-s) \int X
\cdot \pa_y } [\del X^{\nu}, 
X^{\bet}]g_{\mu\nu}(y)g_{\alp\bet}(y)~.
\end{eqnarray}
Just as for the RR-case, the Gauss-Law,
\begin{eqnarray}
  \label{eq:39}
  \int_0^1 ds [X^{\mu}, e^{s\int X \cdot \pa_y} \dot{X}^{\nu}
  e^{(1-s)\int X \cdot \pa_y}] g_{\mu\nu}(y)=0~,
\end{eqnarray}
tells us that fluctuations parallel to the diagonal non-static
background  
are non-dynamical. In addition the fluctuation must be orthogonal to
the 
background velocity; for a $2\times 2$ system in a non-static diagonal
background the orthogonality conditions read (to first approximation in
 a weak background metric) $b \cdot \Del
= 0$ and $b\cdot \dot{\Del}=0$. Dropping crossterms where
$\del X^{\nu}$ is contracted with $X^{\mu}$,   
the remaining physical terms of the potential are
\begin{eqnarray}
  \label{eq:40}
  \del^2 S_{pot} &=&  \lam^2  \int d\tau \int_0^1 ds ~ \tr~e^{s \int X \cdot \pa_y}
[\del X^{\mu},X^{\alp}] e^{(1-s) \int X \cdot \pa_y } [\del X^{\nu},
X^{\bet}] g_{\mu\nu}(y)g_{\alp\bet}(y)~. 
\end{eqnarray}
Substituting the background values and explicit parameterization of
$X^{\mu}$ and $\del X^{\mu}$ for the two by two system from eqs.
(\ref{eq:32}), the ``elementary'' commutator is the same as in
\eqref{eq:8}, 
\begin{eqnarray}
  \label{eq:41}
  [\del X^{\mu},X^{\alp}] = \left( \mtrix{0 & -b^{\mu}\Del^\alp \cr
  \bar{b}^{\mu}\Del^{\alp} & 0 } \right)~.
\end{eqnarray}
The exponentials are again diagonal and the trace can be evaluated as before. We get
\begin{eqnarray}
  \label{eq:42}
 && \hspace{-.3in}
\del^2 S_{pot} =  \lam^2  \int d\tau \int_0^1 ds ~ \left
  [ -f_1(s)f_2(1-s) 
    b^{\mu}\bar{b}^{\nu}- f_2(s)f_1(1-s)b^{\nu}\bar{b}^{\mu} \right]
\Del^{\alp}\Del^{\bet} g_{\mu\nu}(y)g_{\alp\bet}(y) \\ \nonumber 
&&\hspace{.2in}=
 \lam^2 \int d\tau \int_0^1 ds ~
\left[- g_{\mu\nu}(x_2 +
    s\Del)g_{\alp\bet}(x_2+s\Del) -
    g_{\mu\nu}(x_1-s\Del)g_{\alp\bet}(x_1-s\Del)\right] b^{\mu}\bar{b}^{\nu}\Del^{\alp}\Del^{\bet}~.
\end{eqnarray}
The final task is to add eqs (\ref{eq:42}) and
  (\ref{eq:36}), diagonalize the mass matrix and inspect whether one
  of the eigenvalues is negative. If so, this signals the presence of
  a gravitational Myers effect.

\subsubsection{Riemann Normal Coordinates}

These results, eqs. \eqref{eq:36} and \eqref{eq:42}, are valid in {\em
  any} coordinate system to linear order in the metric. 
As we mentioned repeatedly, this is due to
  the use of the weak gravity approximation. To extend beyond linear
  order, while maintaining the determining characteristic of a theory
  coupled to gravity, diffeomorphism invariance, we found that it is
  most readily implemented in an {\em indirect} way as a base-point
  independence constraint between Riemann normal coordinate systems
  \cite{DeBoer:2001uk}. Preferential coordinates in a coordinate
  invariant theory may sound odd, but it is useful to
  think of the noncommuting structure as defined in the tangent space
  (to a (base)point $\cP$). And normal coordinates 
  are the coordinates of
  the tangent space at $\cP$ pulled back to the base manifold. 
  This is also the way one generalizes
  standard non-commutative geometry, i.e. with central
  non-commutativity $[x^i,x^j]=\th^{ij},~[x^i,\th^{jk}]=0$, to curved
  Poisson manifolds \cite{cataneo}.

Because RNC systems are relevant in this sense to non-abelian D0-brane
dynamics, let us also compute the leading contribution to masses of 
off-diagonal fluctuations around geodesics explicitly in such a
coordinate system. Performing geodesic computations in RNC systems has
the additional benefit that one has in effect already solved the
geodesic equation. By definition geodesics through the basepoint $\cP$
are ``straight'' lines, linear in the affine parameter. Again the price
to pay is one of range of validity. RNC are valid in patches
  where geodesics through $\cP$ do not cross. In practice, 
normal coordinates are useful 
in the patch around $\cP$ as long as the geodesic Taylor 
expansion converges. 
In the Riemann normal coordinate system itself this translates into
the fact that one may express the metric and any other tensor 
as a Taylor series around
$\cP$, the origin of the RNC system (see
e.g. \cite{three,mukhi,Alvarez-Gaume:hn}).  
This is how RNC systems are generally used: as approximations
near $\cP$. Taylor expanding the action itself,
the first correction in distance to the origin occurs at fourth order
(by construction) and is proportional to the Riemann tensor. We will
make this approximation to fourth order also for the
potential. At this level the weak gravity approximation is  
academic, as in Riemann normal coordinates with Taylor expanded
metric, the weak gravity approximation is the same as an expansion in
curvatures; i.e. we keep derivatives of the Riemann tensor, but discard
higher powers. 

To illustrate the advantage of RNC, let us start again from the
action. As eq. \eqref{eq:mass} will show the result will be the
same as that obtained from substituting RNC in eqs \eqref{eq:36} and
\eqref{eq:42}.

Ignoring the kinetic term, the action in RNC around a point $p$ and to fourth order is
\cite{DeBoer:2001uk},\footnote{For higher order terms, it is probably
  more convenient to use the 
auxiliary $dN^2$ dimensional tensors defined
  in \cite{DeBoer:2001uk}.} 
\begin{equation}
    S = \frac{1}{2}\int d\tau ~ \frac{m}{3}
    {R}_{\mu\alp\bet\nu}(p)\str(X^{\alp}X^{\bet}{\dot
    X}^{\mu}{\dot X}^{\nu})+ 
    \frac{\lambda^2}{2}g_{\mu\nu}(p)
    g_{\alp\bet}(p)\tr([X^{\mu},X^{\alp}][X^{\nu},X^{\bet}]).    
\end{equation}
Note that the Riemann tensor and the metric evaluated at the
base-point $p$ are well-defined
``classical'' abelian objects. It is only the deviations from the
basepoint which are promoted to $U(N)$ matrices.

Computing the second order variation of the action in the
fluctuation $\delta X$, we find
\begin{equation}
\begin{split}
    \delta^2 S =  \frac{1}{2} \int d\tau &\left[ 
2 m {R}_{\bet\alp\mu\nu}(p)\str(\delta X^{\alp} \delta X^{\mu}{\dot
  X}^{\bet}{\dot X}^{\nu})\right. \\  
&+ 2\lambda^2 g_{\mu\nu}(p)g_{\alp\bet}(p) \tr(\delta X^{\mu} [\delta X^{\alp},[X^{\nu},X^{\bet}]]
)\\ 
&+ 2\lambda^2 g_{\mu\nu}(p)g_{\alp\bet}(p) \tr(\delta X^{\mu} [X^{\alp},[\delta X^{\nu},X^{\bet}]]
)\\ 
&\left. + 2\lambda^2 g_{\mu\nu}(p)g_{\alp\bet}(p) \tr(\delta X^{\mu} [X^{\alp},[X^{\nu},\delta X^{\bet}]] )
\right]. 
    \label{eq:quadrfluc}
\end{split}
\end{equation}
Specializing to the case
of two D0-branes with
\begin{eqnarray}
    \delta X^{\mu} &=& \left ( \begin{array}{cc} 0 & b^{\mu} \\
    \bar{b}^{\mu} & 0 
    \end{array} \right), \non
    X^{\mu} &=& \left ( \begin{array}{cc} x_1^{\mu} & 0 \\ 0 &
    x_2^{\mu} \end{array} 
    \right),  
\end{eqnarray}
imposing the Gauss law constraint, and recalling the
definitions $\Del^{\mu}=x_1^{\mu}-x_2^{\mu},~\bar{x}^{\mu}\equiv \hlf( 
x_1^{\mu}+x_2^{\mu})$,  
we finally arrive at the expression for the mass
matrix 
\begin{eqnarray}
\label{eq:mass}
    \delta^2 S &=& - \int d\tau \left[  b^{\nu}  ( \right.2\lambda^2
    g_{\mu\nu}(p)g_{\alp\bet}(p)\Delta^{\alp}\Delta^{\bet}  \\
    \nonumber
&& \hspace{.5in}
\left.- m({R}_{\alp\mu\nu\bet}(p)+{R}_{\alp\nu\mu\bet}(p))({\dot{\bar{x}}}^{\alp}{\dot{\bar{x}}}^{\bet} + \tfrac{1}{12} {\dot\Delta}^{\alp}{\dot\Delta}^{\bet})){\bar
    b}^{\mu}\right]  \\ \nonumber
&\equiv& - \int d\tau b^{\nu} m_{\nu\mu}(p) {\bar
    b}^{\mu}  ~.
\end{eqnarray}
This is indeed nothing but eqs. \eqref{eq:36} and \eqref{eq:42}
combined after the substitution $\cR_{\mu\nu}({x})=
R_{\mu\alp\bet\nu}(p)\dot{x}^\alp\dot{x}^{\bet} +\cO(x^3)$.  This
computation thus justifies our neglect of the connection terms in the kinetic
operator to the mass-matrix of the fluctuations.

%%%%%%%%%%%%%%%%%%%%%%%%%%%%%%%%%%%%%%%%%%%%%%%%%%
\subsection{D0-branes on a sphere.}
%%%%%%%%%%%%%%%%%%%%%%%%%%%%%%%%%%%%%%%%%%%%%%%%%%

The equivalence between quadratic fluctuations and the geodesic
deviation equation predicts that a tachyonic mode could only be
present in negatively curved region of space-time. Here and in the
next subsection 
we will confirm this prediction by computing the
mass-matrix of quadratic fluctuations on the canonical examples of
positively and negatively curved spacetimes: respectively 
the sphere and the hyperboloid.

Consider $N$ D0-branes on $\RR \times S^2$ and ignore the
remaining seven directions. For the two-sphere we chose the metric
\begin{eqnarray}
  \label{eq:3}
  ds^2 &=&-d\tau^2+
  \left(\frac{d\rho^2}{1-\rho^2/D^2}\right)+\rho^2d\phi^2 \non
  &=& -d\tau^2+dx^2+dy^2 +
  (dx+dy)^2\left(\frac{(x^2+y^2)}{D^2-(x^2+y^2)}\right) ~.
\end{eqnarray}
The weak gravity approximation is applicable for $\rho \ll D$, the
radius of the two-sphere. 

As background configuration we will consider
for simplicity two D0-branes on geodesics approximately 
through the origin $\rho=0$. That is, we choose 
$\dot{x}^{\phi}_{\ell} =0,~(\ell=1,2)$ for both D0-branes. To 
explicitly satisfy
the orthogonality conditions implied by the Gauss law, we set
 the angular separation of the D0-branes $\Del^{\phi}$ to zero:
 $\Del^{\phi}=0$. This ensures that the fluctuation $b^\phi$ is
 unconstrained. If the D0-branes are truly coincident
in the seven additional
dimensions as well, i.e. $\Del^i=0$, $i=1\ldots 7$, 
the Gauss law constrains the $b^r$
fluctuation to vanish.\footnote{This is a completely arbitrary choice. 
The Gauss law orthogonality
  conditions must hold in the full nine transverse dimensions. One has
  a lot of freedom in choosing how to satisfy them. E.g. separating
  the D0-branes also slightly in one of the seven additional
  dimensions forces a linear combination of the $b^i$ fluctuations to
  vanish.} The situation we therefore consider is one with two
D0-branes moving on the same great arc through the origin, but slightly
separated along the direction of the arc.

To compute the contribution from the kinetic term to the mass-matrix
we need to evaluate the intermediate quantity
$\cR_{\mu\nu}(x)\equiv 
  R_{\mu\alp\bet\nu}(x)\dot{x}^{\alp}\dot{x}^{\bet} $ 
defined below eq. \eqref{eq:31}. The Riemann tensor for a sphere is
well known and using that {\em only} $\dot{x}^{\rho} \neq 0$ we find
a single nonzero component\footnote{In our conventions the Riemann
  tensor of an $n$-sphere equals $R_{\mu\nu\rho\tau}= D^{-2}(g_{\mu\rho}g_{\nu\tau}-g_{\nu\rho}g_{\mu\tau})$.} 
\begin{eqnarray}
  \label{eq:4}
  \cR_{\phi\phi}(x) &=& R_{\phi\rho\rho\phi}(\dot{x}^{\rho})^2 \non
&=& -\ove{D^2}\left[g_{\phi\phi}(x)(\dot{x}^{\rho})^2\right] =
-\ove{D^2}(x^{\rho})^2(\dot{x}^{\rho})^2 ~.
\end{eqnarray}
The mass-matrix contribution from the kinetic term is therefore simply
\begin{eqnarray}
  \label{eq:51}
  \del^2 S_{kin,m} =  \int d\tau \int_0^1ds~ b^{\phi}\bar{b}^{\phi}
  \left[ - \frac{m}{D^2}
  \left(g_{\phi\phi}(x_2+s\Del)(\dot{x}^{\rho}_2+s\dot{\Del}^{\rho})^2
  +
  g_{\phi\phi}(x_1-s\Del)
  (\dot{x}^{\rho}_1-s\dot{\Del}^{\rho})^2\right)\right]   ~.
\end{eqnarray}
Note that the unphysical fluctuation $b^\rho$ does not couple to the
background curvature.

Making the additional approximation that $\dot{\Del}^{\rho}$ is 
negligible
(the D0-branes 
have approximately the same initial speed away from the origin; 
their motion
only differs in the $\phi$-direction and a small separation in 
$\rho$), we find
($\bar{x}^{\rho}=\hlf(x_1^{\rho}+x_2^{\rho})$) 
\begin{eqnarray}
  \label{eq:53}
  \del^2 S_{kin} =  \int d\tau~ b^{\phi}\bar{b}^{\phi}
  \left[ - \frac{8m}{D^2} \left( 
  (\bar{x}^{\rho})^2+\ove{12}(\Del^{\rho})^2 \right)
  (\dot{\bar{x}}^{\rho})^2 \right]~.
\end{eqnarray}

The potential term is diagonal and yields
\begin{eqnarray}
  \label{eq:5}
  \del^2S_{pot} &=& \lam^2\int d\tau \int_0^1 ds~
  b^{\phi}\bar{b}^{\phi} \left[-8(
  \bar{x}^{\rho})^2\Del^2-4(\bar{x}^{\rho})^2(h_{\rho\rho}(x_1)+h_{\rho\rho}(x_2))
  \Del^{\rho}\Del^{\rho}  
  + \cO(\Del^4)\right] \non
&&\hspace{1in}+ b^{\rho}\bar{b}^{\rho}\left[-2\eta_{\rho\rho}\Del^2 -
  2h_{\rho\rho}(\bar{x})(\Del^2+\eta_{\rho\rho}
  (\Del^{\rho})^2)+\cO(\Del^4)\right]   
\end{eqnarray}
The unphysical 
$b^{\rho}$ sector is therefore the same as for the static
situation. The masses of the $b^{\rho}$ fluctuations are purely 
determined by D-geometry and will be positive.  
The physical $b^{\phi}$-sector, however, is
affected. Adding the two terms we find 
the mass for the off-diagonal fluctuation
$b^{\phi}$ to be 
\begin{eqnarray}
  \label{eq:50}
  m^2_{b^{\phi}}&=&  \frac{8m}{D^2}\left[{(\bar{x}^{\rho})^2} + \ove{12}
(\Del^\rho)^2\right](\dot{\bar{x}}^\rho)^2 +8\lam^2\left[
(\bar{x}^{\rho})^2\Del^2+
(\bar{x}^{\rho})^2h_{\rho\rho}(\bar{x})(\Del^{\rho})^2  
\right] 
+\cO(\Del^4)
\end{eqnarray}
which is positive semidefinite. On a sphere, as expected, and by rough
generalization on any positively curved patch of 
spacetime, no gravitationally
induced Myers effect occurs.

\subsubsection{RNC}

For those metrics where one knows a Riemann normal coordinate system
the computation is much cleaner and more perspicacious. 
To illustrate its
benefits let us repeat the sphere computation starting from RNC.
Since the sphere is a homogeneous space the transformation from the
standard metric to RNC is easily found and one obtains the
metric: 
\begin{equation}
\label{eq:56}
    \dif s^2_{S^2, RNC} = \dif z^2_x + \dif z^2_y
    -\frac{1}{3D^2}(z_x\dif z_y - z_y\dif z_x)^2 +
    \mathcal{O}(z^3). 
\end{equation}
One easily checks that geodesics through the origin $z^i=0$ can be
written as $z^i = v^i\tau$ as required. Furthermore one immediately
reads off the Riemann tensor at the origin $z^i=0$:
\begin{equation}
\label{eq:55}
  ds^2_{RNC} =
  \eta_{\mu\nu}dx^{\mu}dx^{\nu}+\ove{3}R_{\mu\alp\bet\nu}(0)
  x^{\alp}x^{\bet}dx^{\mu}dx^{\nu}+\ldots~~~~  
   \Rightarrow~~~~ R_{xyyx}(0) = -\frac{1}{D^2}.
\end{equation}
The two D0-brane system we are considering has both moving along
the same great arc through the origin, but slightly separated along
this arc. If we choose the $z^x$-axis as this arc, this implies that
$\dot{z}_i^y=0$, $\Del^y=0$ and $b^x$ is the unphysical fluctuation.

Substituting this data into equation \eqref{eq:mass}, we find:
\begin{eqnarray}
\label{eq:54}
    \del^2S &=& - \int d\tau ~ \left[ b_x 
\left(  
2\lam^2\Del^2
\right)
\bar{b}_x \right.
\non    &&\left.
\hspace{0.55in} + b_y\left( 
2\lam^2\Del^2
+\frac{2m}{D^2}\left((\dot{\bar{z}}^x)^2+
\ove{12}(\dot{\Del}^x)^2\right)
\right ) \bar{b}_y\right]~.
\end{eqnarray}
Again the mass-matrix for the unphysical fluctuation $b^x$ is that
given by D-geometry and strictly positive definite. For the physical
$b^y$ fluctuation 
the mass matrix is explicitly positive semidefinite, and we
see that there is no tachyon in this case. The physical picture
explains this in simple terms. There are two forces acting: 
the force due to the geodesic
deviation and the force of open string between the branes. In this
positive curvature background the geodesic deviation works in the same
direction as the force by the stretched open strings, leading to a
stable configuration.

%%%%%%%%%%%%%%%%%%%%%%%%%%%%%%%%%%%%%%%%%%%%%%%%%%
\subsection{D0-branes on a hyperboloid}
%%%%%%%%%%%%%%%%%%%%%%%%%%%%%%%%%%%%%%%%%%%%%%%%%%
\label{sec:d0-bran-hyperb}

So far we confirmed our supposition that on positively curved patches
standard 
``individual'' geodesic behaviour is a stable solution. 
The real question, however, is whether negatively curved
patches do destabilize this solution and therefore hint towards the
existence of a gravitational Myers effect. That this is indeed
the case, is now easily seen. For two nearly coincident D0-branes on 
a two-dimensional 
hyperboloid, the mass-matrix of fluctuations is simply
eq. \eqref{eq:54} with the substitution $D^2 \rar -D^2$. The
gravitational tidal force has changed sign and can now counterbalance
the attractive string potential. Specifically there is a tachyonic
instability in the spectrum iff the D-brane separation is small
compared to the velocity times the curvature:
\begin{eqnarray}
  \label{eq:57}
  2\lam^2 \Del^2\, \ll\,
  \frac{2m}{D^2}\left((\dot{\bar{z}}^x)^2+\ove{12}(\dot{\Del}^x)^2\right)~, 
\end{eqnarray}
Note, however, that for large separations
the attractive force of the open strings dominates the
repelling force of the background geometry, and the tachyon
disappears. 

To truly test whether we have found a tachyonic instability one
question remains. The tachyon appears to be evident for large speeds
or small separations. Both extremes, however are outside the range of
validity of the action (eqs. \eqref{eq:3a}-\eqref{eq:3c} and
\eqref{eq:47}). Large speeds violate the
truncation to second order in derivatives, eq. \eqref{eq:3b}, while
separations $\Del$ must be larger
than the eleven-dimensional Planck length, eq. \eqref{eq:3c}.
We need to check whether the
inequality \eqref{eq:57}, signaling the instability, 
can be satisfied within these
bounds. Without loss of generality, we may simplify the inequality by
approximating 
$\dot{\Del}^x \simeq 0$. 
Substituting $\dot{z}^x=v$ with $|v| \ll 1$, expressing $m$ and
$\lam$ in string units $g_s$, $\ell_s$ and 
multiplying both sides of \eqref{eq:57} 
by $\ell_s^2$, we obtain
\begin{eqnarray}
  \label{eq:58}
  \frac{\Del^2}{\ell_s^2} \ll v^2 \frac{\ell_s^2}{D^2}
   \ll 1~.
\end{eqnarray}
In words, the condition for a tachyon to be present is
that the separation in string units must be less then the velocity
times the curvature in string units. We have added the implicit
weak gravity and Born-Infeld condition: both sides must be less
than unity. Most importantly, however, the separation
$\Del$ must also be larger than the eleven dimensional
Planck length, eq. \eqref{eq:3c}:
\begin{eqnarray}
  \label{eq:52}
  g_s^{1/3}\ell_s \ll {\Del^i} \ll 
\ell_s  ~.
\end{eqnarray}
The window where eq. \eqref{eq:58} can be satisfied
within the range of eq. \eqref{eq:52} is when the velocity times the
curvature is much larger than the eleven dimensional Planck length:
\begin{eqnarray}
  \label{eq:59}
 g_s^{2/3} \ll v^2 \frac{\ell_s^2}{D^2}~.
\end{eqnarray}
This is easily satisfied for very weak string coupling.
The region where the ``individual'' geodesic solution is unstable
thus falls easily within the region of validity of the action.
This
should convince us that the existence of a lesser energetic
collective mode is highly plausible. It appears that a purely
gravitational Myers effect exists.

%%%%%%%%%%%%%%%%%%%%%%%%%%%%%%%%%%%%%%%%%%%%%%%%%%%%%%%%%%%%%%%%%%%%%%

\subsection{D0-branes near Schwarzschild black holes}

The sphere and the hyperboloid are abstract situations, that are
unlikely to occur in nature and do not confront fundamental questions
in quantum gravity. Can a gravitational Myers effect occur in
physically interesting situations? As primary physical 
testbed we will study
the Schwarzschild black hole. In the weak gravity approximation the
$d$-dimensional Schwarzschild metric
may be approximated by 
\begin{eqnarray}
  \label{eq:23}
  ds^2 &=& -\left(1-\frac{2M}{r^{d-3}}\right) dt^2 +
  \left(1-\frac{2M}{r^{d-3}}\right)\inv dr^2 + r^2 d\Ome^2_{d-2} \\
&=&
  \eta_{\mu\nu} dx^{\mu}dx^{\nu} + \frac{2M}{r^{d-3}} (dt^2+dr^2) +
  \cO(M^2) ~.
\end{eqnarray}
The Schwarzschild spacetime confronts us with an 
extremely important fact. Generically 
the ambient spacetime will be curved
in the timelike/null directions that are not ``orthogonal'' 
to the D0-brane. So far, we have 
implicitly {\em i)} discussed metrics which only have curvature in the
directions {\em transverse} to the D0-brane and {\em ii)} 
always chosen the
physical gauge for the time-like direction.  How do we 
deal with non-abelian D0-branes in 
metrics with non-transverse curvature? 
In a separate article \cite{ourselves} 
we will show that
diffeomorphism invariance 
allows one to simply extend
the range of the index of the fields to {\em include} the direction
parallel to the worldvolume of the D0-brane  {\em on the
  condition that one chooses the physical gauge $X^0= \tau \cdot
  \one$ at the end}. This is what one intuitively
expects.  It also has an important
consequence for the computation of the mass-matrix of quadratic
fluctuations. As the standard metric choice contains only deviations
in the radial and timelike directions, only the off-diagonal modes
$b^r$ and $b^t$ could potentially be tachyonic. The physical gauge
choice, however, tells us that $b^t$ is pure gauge and
unphysical. The mass matrix therefore has only a single interesting 
entry, that for the off-diagonal mode $b^r$. To ensure that the Gauss
law orthogonality constraints do not affect this mode we choose the
zero radial separation between the two D0-branes: $\Del^{\rho}=0$. The other off-diagonal
modes $b^i$ have
their standard geodesic-distance masses determined by D-geometry
subject to the Gauss law, and
we may ignore them.

Collecting then
the various pieces we find explicitly 
\begin{eqnarray}
  \label{eq:43}
 \del^2 S_{kin,r} + \del^2 S_{pot,r}  &=& \int d\tau \int_0^1 ds
 ~b^r \bar{b}^r\left[ m\cR_{rr} (x_2+s\Del) +m\cR_{rr}(x_1-s\Del)
  \right. \non
&&\left. -
 \eta_{rr}2\lam^2
 \Del^2\left(1+h(x_2+s\Del)+h(x_1-s\Del)\right)\right]~,  
\end{eqnarray}
where we have used
\begin{eqnarray}
h_{rr}=h_{tt} &\equiv & h(r) = \frac{2M}{r^{d-3}}~, \non \cR_{rr} &=&
R_{rttr}\dot{x}^t \dot{x}^t \non
\label{eq:44}
&=& R_{rttr} =\frac{M (d-3)(d-2)}{r^{d-1}}\left(1 +\cO
\left(\frac{M}{r^{d-3}}\right)\right)~.
\end{eqnarray}
Note that $h(r)$ and $\cR_{rr}(r)$ are functions only of the radial
direction $r$, and that we have chosen $\Del^r=0$. This 
means that we may
simply replace the arguments $x_2^r+s\Del^r$, $x_1^r-s\Del^r$ of
these functions by the center of 
mass $\bar{r} = \hlf(r_1+r_2)$. The mass matrix therefore
immediately follows from eqs. \eqref{eq:43} and \eqref{eq:44} 
and reads (ignoring higher order corrections)
\begin{eqnarray}
  \label{eq:72}
  \del^2 S 
&=& \int d\tau\, \bar{b}^r \left
  [ \frac{2 m M(d-3)(d-2)}{\bar{r}^{d-1}}
  - 2\lam^2 \Del^2 
  \left(1+\frac{4mM}{\bar{r}^{d-3}}\right)\right] b^r  
%\\ && ~~~~~~~~~~~~~~~~
~.\nonumber
\end{eqnarray}
This corresponds to a negative mass squared iff
\begin{eqnarray}
  \label{eq:73}
2\lam^2 \Del^2
  \left(1+\frac{4mM}{\bar{r}^{d-3}}\right)&\ll&\frac{2mM(d-3)(d-2)}{\bar{r}^{d-1}}
 ~.
\end{eqnarray}
Conform physical intuition, we see that close to the black hole
  the Myers tachyon appears, 
whereas far-away ($\bar{r} \gg 1$), eq. (\ref{eq:72})
  reduces to the
  regular flat space mass-matrix. We do still 
need to test that we remain within the regime of validity
of the action. In the weak gravity approximation we may approximate
  the l.h.s. with the leading term.
Denote the curvature scale, eq. \eqref{eq:44}, with
\begin{eqnarray}
\label{eq:80} 
\frac{(d-3)(d-2)M}{\bar{r}^{d-1}} = \frac{1}{D(\bar{r})^2}~.
\end{eqnarray}
and rewrite all quantities in string units.
We see that we obtain a tachyonic mode if
\begin{equation}
\label{eq:81} 
\frac{\Del^2}{\ell_s^2} \ll \frac{\ell_s^2}{D(\bar{r})^2}~,
\end{equation}
where again both sides must also be less than unity. As repeatedly emphasized, the
separation must in addition be larger than the eleven-dimensional
Planck length,
\eqref{eq:3c}:
\begin{eqnarray}
  \label{eq:61}
  g_s^{1/3}\ell_s \ll \Del \ll \ell_s ~.
\end{eqnarray}
A tachyon will therefore appear if the curvature in string units is
large compared to 
the string coupling
\begin{eqnarray}
  \label{eq:48}
  g_s^{2/3} \ll \frac{\ell_s^2}{D(\bar{r})^2}
\end{eqnarray}
The Schwarzschild curvature
potentially induces near-horizon 
collective Myers behaviour in D0-branes! 
Note the close similarity with the
condition we found for the hyperboloid, eq. \eqref{eq:59}. This makes
the claim that the qualitative nature of the local curvature, positive
or negative, determines the existence of a tachyonic mode in the
mass-matrix of off-diagonal fluctuations. Locally 
negative curvature patches of spacetime potentially 
induce a geometric Myers effect.

\section{Conclusion}

We have provided evidence that $N$
nearly superposed D0-branes following independent geodesic motion are
unstable in a patch of negatively curved spacetime. This conclusion
follows from a fluctuation analysis of off-diagonal modes around
the diagonal solution which has the geometric interpretation of
computing {\em non-abelian geodesic deviation.} Given this evidence,
one is prompted to ask if and what the collective
configuration is that the D0-branes fall into. This final non-abelian 
configuration
will 
depend heavily on the specific properties of the spacetime, and is
therefore difficult to compute in general, although some specific
highly symmetric cases may be solvable.  
In the RR-Myers effect 
the properties of the
Ramond-Ramond gauge field provide a handle on the 
interpretation of the collective lesser energetic configuration as dipole
configurations of fundamental sources. Everything
couples to gravity, however, and this makes a gravitational Myers
effect less transparent and more universal at the same
time. Generically supersymmetry will also be absent, which makes it
even less 
clear whether the resultant configuration is a identifiable 
polarized cloud of
specific D-branes and/or strings. On the other hand, the indication
that a gravitationally induced Myers effect may occur near the horizon
of a Schwarzschild black hole hints at a universal aspect, which
through the link between D0-brane mechanics and M-theory could
possibly have consequences for our understanding of GR-singularities
in string theory. However, we will need to understand the non-commutative non-abelian geometry underlying
D0-brane mechanics better to make progress in this direction. One
avenue which a possible near-horizon Myers effect suggests to explore
is a connection between
D0-brane mechanics
and the non-commutative shockwave-black hole geometries of 't Hooft \cite{hooft}.

Whether a 
gravitational Myers effect truly exists and whether it may contribute
to 
answering fundamental questions in quantum gravity, however, remain questions for the future. 

\bigskip
{\bf Acknowledgments:} We sincerely wish to thank both the hosts and
participants of the Amsterdam Summer Workshop as well as the Aspen
Center for Physics for the hospitality and enlightening
discussions. EG thanks Marcus Spradlin and
KS thanks Dan Kabat, Gilad Lifschytz and David Lowe for
sharing their insights. JdB and JW thank Stichting FOM for support. 
EG is supported by Frank and Peggy Taplin, and
by NSF grant PHY-0070928.
KS gratefully acknowledges support from DOE grant
DE-FG-02-92ER40699.

%%%%%%%%%%%%%%%%%%%%%%%%%%%%%%%%%%%%%%%%%%%%%%%

\end{document}